\documentclass[11pt]{article}
\pdfoutput=1
\usepackage{graphicx}
\usepackage{amssymb}
\usepackage{amsmath}
\usepackage[left]{lineno}
\usepackage{booktabs}
\usepackage{multirow}
\usepackage[title]{appendix}
\usepackage{lineno}
\usepackage{xcolor} 
\usepackage{subcaption} 
\usepackage{cite} 
\usepackage{xspace}
\usepackage[normalem]{ulem}

\def\mmis{m^2_\mathrm{miss}}
\def\Kpnn{K^{+}\rightarrow\pi^{+}\nu\bar{\nu}}
\def\KpiX{K^{+}\rightarrow\pi^{+}X}
\def\geant{\mbox{\textsc{Geant4}}\xspace}

\begin{document}
\pagenumbering{arabic}
\centerline{\LARGE EUROPEAN ORGANIZATION FOR NUCLEAR RESEARCH}
\vspace{15mm}
\begin{flushright}
 \vspace{2mm}
\end{flushright}
\vspace{15mm}

\begin{center}
\Large{\bf Search for a feebly interacting particle {\boldmath $X$} \\in the decay {\boldmath $K^{+}\rightarrow\pi^{+}X$} \\
\vspace{5mm}
}
The NA62 Collaboration
\end{center}
\begin{abstract}
A search for the $\KpiX$ decay, where $X$ is a long-lived feebly interacting particle, is performed through an interpretation of the $\Kpnn$ analysis of data collected in 2017 by the NA62 experiment at CERN.
Two ranges of $X$ masses, $0$--$110\,\text{MeV}/c^{2}$ and $154$--$260\,\text{MeV}/c^{2}$, and lifetimes above $100\,\text{ps}$ are considered.
The limits set on the branching ratio, $\text{BR}(K^{+}\rightarrow\pi^{+}X)$, are competitive with previously reported searches in the first mass range, and improve on current limits in the second mass range by more than an order of magnitude.
\end{abstract}
\vspace{20mm}

\begin{center}
{\em Accepted for publication by JHEP} 
\end{center}

\clearpage
\begin{center}
{\Large The NA62 Collaboration$\,$\renewcommand{\thefootnote}{\fnsymbol{footnote}}%
\footnotemark[1]\renewcommand{\thefootnote}{\arabic{footnote}}}\\
\end{center}
 \vspace{3mm}

\begin{raggedright}
\noindent
{\bf Universit\'e Catholique de Louvain, Louvain-La-Neuve, Belgium}\\
 E.~Cortina Gil,
 A.~Kleimenova,
 E.~Minucci$\,$\footnotemark[1]$^,\,$\footnotemark[2],
 S.~Padolski$\,$\footnotemark[3],
 P.~Petrov,
 A.~Shaikhiev$\,$\footnotemark[4],
 R.~Volpe$\renewcommand{\thefootnote}{\fnsymbol{footnote}}\footnotemark[1]\renewcommand{\thefootnote}{\arabic{footnote}}^,$\,\footnotemark[5]\\[2mm]

{\bf TRIUMF, Vancouver, British Columbia, Canada}\\
 T.~Numao,
 B.~Velghe\\[2mm]

{\bf University of British Columbia, Vancouver, British Columbia, Canada}\\
 D.~Bryman$\,$\footnotemark[6],
 J.~Fu$\,$\footnotemark[7]\\[2mm]

{\bf Charles University, Prague, Czech Republic}\\
 T.~Husek$\,$\footnotemark[8],
 J.~Jerhot$\,$\footnotemark[9],
 K.~Kampf,
 M.~Zamkovsky\\[2mm]

{\bf Institut f\"ur Physik and PRISMA Cluster of excellence, Universit\"at Mainz, Mainz, Germany}\\
 R.~Aliberti$\,$\footnotemark[10],
 G.~Khoriauli$\,$\footnotemark[11],
 J.~Kunze,
 D.~Lomidze$\,$\footnotemark[12],
 R.~Marchevski$\,$\footnotemark[13],
 L.~Peruzzo,
 M.~Vormstein,
 R.~Wanke\\[2mm]

{\bf Dipartimento di Fisica e Scienze della Terra dell'Universit\`a e INFN, Sezione di Ferrara, Ferrara, Italy}\\
 P.~Dalpiaz,
 M.~Fiorini,
 I.~Neri,
 A.~Norton,
 F.~Petrucci,
 H.~Wahl\\[2mm]

{\bf INFN, Sezione di Ferrara, Ferrara, Italy}\\
 A.~Cotta Ramusino,
 A.~Gianoli\\[2mm]

{\bf Dipartimento di Fisica e Astronomia dell'Universit\`a e INFN, Sezione di Firenze, Sesto Fiorentino, Italy}\\
 E.~Iacopini,
 G.~Latino,
 M.~Lenti,
 A.~Parenti\\[2mm]

{\bf INFN, Sezione di Firenze, Sesto Fiorentino, Italy}\\
 A.~Bizzeti$\,$\footnotemark[14],
 F.~Bucci\\[2mm]

{\bf Laboratori Nazionali di Frascati, Frascati, Italy}\\
 A.~Antonelli,
 G.~Georgiev$\,$\footnotemark[15],
 V.~Kozhuharov$\,$\footnotemark[15],
 G.~Lanfranchi,
 S.~Martellotti,
 M.~Moulson,
 T.~Spadaro\\[2mm]

{\bf Dipartimento di Fisica ``Ettore Pancini'' e INFN, Sezione di Napoli, Napoli, Italy}\\
 F.~Ambrosino,
 T.~Capussela,
 M.~Corvino$\,$\footnotemark[13],
 D.~Di Filippo,
 P.~Massarotti,
 M.~Mirra,
 M.~Napolitano,
 G.~Saracino\\[2mm]

{\bf Dipartimento di Fisica e Geologia dell'Universit\`a e INFN, Sezione di Perugia, Perugia, Italy}\\
 G.~Anzivino,
 F.~Brizioli,
 E.~Imbergamo,
 R.~Lollini,
 R.~Piandani$\,$\footnotemark[16],
 C.~Santoni\\[2mm]

{\bf INFN, Sezione di Perugia, Perugia, Italy}\\
 M.~Barbanera$\,$\footnotemark[17],
 P.~Cenci,
 B.~Checcucci,
 P.~Lubrano,
 M.~Lupi$\,$\footnotemark[18],
 M.~Pepe,
 M.~Piccini\\[2mm]

{\bf Dipartimento di Fisica dell'Universit\`a e INFN, Sezione di Pisa, Pisa, Italy}\\
 F.~Costantini,
 L.~Di Lella,
 N.~Doble,
 M.~Giorgi,
 S.~Giudici,
 G.~Lamanna,
 E.~Lari,
 E.~Pedreschi,
 M.~Sozzi\\[2mm]

{\bf INFN, Sezione di Pisa, Pisa, Italy}\\
 C.~Cerri,
 R.~Fantechi,
 L.~Pontisso,
 F.~Spinella\\[2mm]
\newpage
{\bf Scuola Normale Superiore e INFN, Sezione di Pisa, Pisa, Italy}\\
 I.~Mannelli\\[2mm]

{\bf Dipartimento di Fisica, Sapienza Universit\`a di Roma e INFN, Sezione di Roma I, Roma, Italy}\\
 G.~D'Agostini,
 M.~Raggi\\[2mm]

{\bf INFN, Sezione di Roma I, Roma, Italy}\\
 A.~Biagioni,
 E.~Leonardi,
 A.~Lonardo,
 P.~Valente,
 P.~Vicini\\[2mm]

{\bf INFN, Sezione di Roma Tor Vergata, Roma, Italy}\\
 R.~Ammendola,
 V.~Bonaiuto$\,$\footnotemark[19],
 A.~Fucci,
 A.~Salamon,
 F.~Sargeni$\,$\footnotemark[20]\\[2mm]

{\bf Dipartimento di Fisica dell'Universit\`a e INFN, Sezione di Torino, Torino, Italy}\\
 R.~Arcidiacono$\,$\footnotemark[21],
 B.~Bloch-Devaux,
 M.~Boretto$\,$\footnotemark[13],
 E.~Menichetti,
 E.~Migliore,
 D.~Soldi\\[2mm]

{\bf INFN, Sezione di Torino, Torino, Italy}\\
 C.~Biino,
 A.~Filippi,
 F.~Marchetto\\[2mm]

{\bf Instituto de F\'isica, Universidad Aut\'onoma de San Luis Potos\'i, San Luis Potos\'i, Mexico}\\
 J.~Engelfried,
 N.~Estrada-Tristan$\,$\footnotemark[22]\\[2mm]

{\bf Horia Hulubei National Institute of Physics for R\&D in Physics and Nuclear Engineering, Bucharest-Magurele, Romania}\\
 A. M.~Bragadireanu,
 S. A.~Ghinescu,
 O. E.~Hutanu\\[2mm]

{\bf Joint Institute for Nuclear Research, Dubna, Russia}\\
 A.~Baeva,
 D.~Baigarashev,
 D.~Emelyanov,
 T.~Enik,
 V.~Falaleev,
 V.~Kekelidze,
 A.~Korotkova,
 L.~Litov$\,$\footnotemark[15],
 D.~Madigozhin,
 M.~Misheva$\,$\footnotemark[23],
 N.~Molokanova,
 S.~Movchan,
 I.~Polenkevich,
 Yu.~Potrebenikov,
 S.~Shkarovskiy,
 A.~Zinchenko$\,$\renewcommand{\thefootnote}{\fnsymbol{footnote}}\footnotemark[2]\renewcommand{\thefootnote}{\arabic{footnote}}\\[2mm]

{\bf Institute for Nuclear Research of the Russian Academy of Sciences, Moscow, Russia}\\
 S.~Fedotov,
 E.~Gushchin,
 A.~Khotyantsev,
 Y.~Kudenko$\,$\footnotemark[24],
 V.~Kurochka,
 M.~Medvedeva,
 A.~Mefodev\\[2mm]

{\bf Institute for High Energy Physics - State Research Center of Russian Federation, Protvino, Russia}\\
 S.~Kholodenko,
 V.~Kurshetsov,
 V.~Obraztsov,
 A.~Ostankov$\,$\renewcommand{\thefootnote}{\fnsymbol{footnote}}\footnotemark[2]\renewcommand{\thefootnote}{\arabic{footnote}},
 V.~Semenov$\,$\renewcommand{\thefootnote}{\fnsymbol{footnote}}\footnotemark[2]\renewcommand{\thefootnote}{\arabic{footnote}},
 V.~Sugonyaev,
 O.~Yushchenko\\[2mm]

{\bf Faculty of Mathematics, Physics and Informatics, Comenius University, Bratislava, Slovakia}\\
 L.~Bician$\,$\footnotemark[13],
 T.~Blazek,
 V.~Cerny,
 Z.~Kucerova\\[2mm]

{\bf CERN,  European Organization for Nuclear Research, Geneva, Switzerland}\\
 J.~Bernhard,
 A.~Ceccucci,
 H.~Danielsson,
 N.~De Simone$\,$\footnotemark[25],
 F.~Duval,
 B.~D\"obrich,
 L.~Federici,
 E.~Gamberini,
 L.~Gatignon,
 R.~Guida,
 F.~Hahn$\,$\renewcommand{\thefootnote}{\fnsymbol{footnote}}\footnotemark[2]\renewcommand{\thefootnote}{\arabic{footnote}},
 E. B.~Holzer,
 B.~Jenninger,
 M.~Koval$\,$\footnotemark[26],
 P.~Laycock$\,$\footnotemark[3],
 G.~Lehmann Miotto,
 P.~Lichard,
 A.~Mapelli,
 K.~Massri,
 M.~Noy,
 V.~Palladino$\,$\footnotemark[27],
 M.~Perrin-Terrin$\,$\footnotemark[28]$^,\,$\footnotemark[29],
 J.~Pinzino$\,$\footnotemark[30]$^,\,$\footnotemark[31],
 V.~Ryjov,
 S.~Schuchmann$\,$\footnotemark[32],
 S.~Venditti\\[2mm]

{\bf University of Birmingham, Birmingham, United Kingdom}\\
 T.~Bache,
 M. B.~Brunetti$\,$\footnotemark[33],
 V.~Duk$\,$\footnotemark[34],
 V.~Fascianelli$\,$\footnotemark[35],
 J. R.~Fry,
 F.~Gonnella,
 E.~Goudzovski,
 L.~Iacobuzio,
 C.~Lazzeroni,
 N.~Lurkin$\,$\footnotemark[9],
 F.~Newson,
 C.~Parkinson$\,$\footnotemark[9],
 A.~Romano,
 A.~Sergi,
 A.~Sturgess,
 J.~Swallow\renewcommand{\thefootnote}{\fnsymbol{footnote}}
\footnotemark[1]\renewcommand{\thefootnote}{\arabic{footnote}}\\[2mm]
\newpage
{\bf University of Bristol, Bristol, United Kingdom}\\
 H.~Heath,
 R.~Page,
 S.~Trilov\\[2mm]

{\bf University of Glasgow, Glasgow, United Kingdom}\\
 B.~Angelucci,
 D.~Britton,
 C.~Graham,
 D.~Protopopescu\\[2mm]

{\bf University of Lancaster, Lancaster, United Kingdom}\\
 J.~Carmignani,
 J. B.~Dainton,
 R. W. L.~Jones,
 G.~Ruggiero\\[2mm]

{\bf University of Liverpool, Liverpool, United Kingdom}\\
 L.~Fulton,
 D.~Hutchcroft,
 E.~Maurice$\,$\footnotemark[36],
 B.~Wrona\\[2mm]

{\bf George Mason University, Fairfax, Virginia, USA}\\
 A.~Conovaloff,
 P.~Cooper,
 D.~Coward$\,$\footnotemark[37],
 P.~Rubin\\[2mm]

\end{raggedright}
%
%
\setcounter{footnote}{0}
\renewcommand{\thefootnote}{\fnsymbol{footnote}}
\footnotetext[1]{Corresponding authors:  J.~Swallow, R.~Volpe, email: joel.christopher.swallow@cern.ch, roberta.volpe@cern.ch}
\footnotetext[2]{Deceased}
\renewcommand{\thefootnote}{\arabic{footnote}}

\footnotetext[1]{Present address: Laboratori Nazionali di Frascati, I-00044 Frascati, Italy}
\footnotetext[2]{Also at CERN,  European Organization for Nuclear Research, CH-1211 Geneva 23, Switzerland}
\footnotetext[3]{Present address: Brookhaven National Laboratory, Upton, NY 11973, USA}
\footnotetext[4]{Also at Institute for Nuclear Research of the Russian Academy of Sciences, 117312 Moscow, Russia}
\footnotetext[5]{Present address: Faculty of Mathematics, Physics and Informatics, Comenius University, 842 48, Bratislava, Slovakia}
\footnotetext[6]{Also at TRIUMF, Vancouver, British Columbia, V6T 2A3, Canada}
\footnotetext[7]{Present address: UCLA Physics and Biology in Medicine, Los Angeles, CA 90095, USA}
\footnotetext[8]{Present address: Department of Astronomy and Theoretical Physics, Lund University, Lund, SE 223-62, Sweden}
\footnotetext[9]{Present address: Universit\'e Catholique de Louvain, B-1348 Louvain-La-Neuve, Belgium}
\footnotetext[10]{Present address: Institut f\"ur Kernphysik and Helmholtz Institute Mainz, Universit\"at Mainz, Mainz, D-55099, Germany}
\footnotetext[11]{Present address: Universit\"at W\"urzburg, D-97070 W\"urzburg, Germany}
\footnotetext[12]{Present address: European XFEL GmbH, D-22761 Hamburg, Germany}
\footnotetext[13]{Present address: CERN,  European Organization for Nuclear Research, CH-1211 Geneva 23, Switzerland}
\footnotetext[14]{Also at Dipartimento di Fisica, Universit\`a di Modena e Reggio Emilia, I-41125 Modena, Italy}
\footnotetext[15]{Also at Faculty of Physics, University of Sofia, BG-1164 Sofia, Bulgaria}
\footnotetext[16]{Present address: Institut f\"ur Experimentelle Teilchenphysik (KIT), D-76131 Karlsruhe, Germany}
\footnotetext[17]{Present address: School of Physics and Astronomy, University of Birmingham, Birmingham, B15 2TT, UK}
\footnotetext[18]{Present address: Institut am Fachbereich Informatik und Mathematik, Goethe Universit\"at, D-60323 Frankfurt am Main, Germany}
\footnotetext[19]{Also at Department of Industrial Engineering, University of Roma Tor Vergata, I-00173 Roma, Italy}
\footnotetext[20]{Also at Department of Electronic Engineering, University of Roma Tor Vergata, I-00173 Roma, Italy}
\footnotetext[21]{Also at Universit\`a degli Studi del Piemonte Orientale, I-13100 Vercelli, Italy}
\footnotetext[22]{Also at Universidad de Guanajuato, Guanajuato, Mexico}
\footnotetext[23]{Present address: Institute of Nuclear Research and Nuclear Energy of Bulgarian Academy of Science (INRNE-BAS), BG-1784 Sofia, Bulgaria}
\footnotetext[24]{Also at National Research Nuclear University (MEPhI), 115409 Moscow and Moscow Institute of Physics and Technology, 141701 Moscow region, Moscow, Russia}
\footnotetext[25]{Present address: DESY, D-15738 Zeuthen, Germany}
\footnotetext[26]{Present address: Charles University, 116 36 Prague 1, Czech Republic}
\footnotetext[27]{Present address: Physics Department, Imperial College London, London, SW7 2BW, UK}
\footnotetext[28]{Present address: Aix Marseille University, CNRS/IN2P3, CPPM, F-13288, Marseille, France}
\footnotetext[29]{Also at Universit\'e Catholique de Louvain, B-1348 Louvain-La-Neuve, Belgium}
\footnotetext[30]{Present address: Department of Physics, University of Toronto, Toronto, Ontario, M5S 1A7, Canada}
\footnotetext[31]{Also at INFN, Sezione di Pisa, I-56100 Pisa, Italy}
\footnotetext[32]{Present address: Institut f\"ur Physik and PRISMA Cluster of excellence, Universit\"at Mainz, D-55099 Mainz, Germany}
\footnotetext[33]{Present address: Department of Physics, University of Warwick, Coventry, CV4 7AL, UK}
\footnotetext[34]{Present address: INFN, Sezione di Perugia, I-06100 Perugia, Italy}
\footnotetext[35]{Present address: Dipartimento di Psicologia, Universit\`a di Roma La Sapienza, I-00185 Roma, Italy}
\footnotetext[36]{Present address: Laboratoire Leprince Ringuet, F-91120 Palaiseau, France}
\footnotetext[37]{Also at SLAC National Accelerator Laboratory, Stanford University, Menlo Park, CA 94025, USA}

\clearpage

\section{Introduction}
\label{sec:intro}
Some scenarios Beyond the Standard Model of particle physics (BSM) include a new light feebly interacting particle $X$, which can be produced in $\KpiX$ decays. In a hidden sector portal framework the new $X$ particle mediates interactions between standard model (SM) and hidden sector fields~\cite{PBC19}. In the Higgs portal scenario, $X$ is a scalar that mixes with the SM Higgs boson; this is realised in inflationary~\cite{Bezrukov10}, scale invariant~\cite{Clarke13}, and relaxion~\cite{Banerjee20} models, which additionally have cosmological implications.
A massless $X$ particle would have the properties of a neutral boson arising from the spontaneous breaking of a global $U(1)$ symmetry~\cite{PDG18}: $X$ may then acquire mass through explicit symmetry breaking. One example, arising from the breaking of a Peccei-Quinn (PQ) symmetry, is an axion~\cite{Weinberg78,Wilczeck82}, which would be a signature of the PQ mechanism and credibly solve the strong CP problem~\cite{PecciQuinn77_1,PecciQuinn77_2}. Such an axion could be flavor non-diagonal~\cite{Hindmarsh99}.
Alternatives, from breaking of the lepton number and flavour symmetries respectively, are majorons~\cite{GelminiRoncadelli81} or familons~\cite{Wilczeck82,DavidsonWali82}.
A QCD axion with mass $\mathcal{O}(10^{-4}\,\text{eV})$ could be a dark matter candidate, and specific axion models can also solve the SM flavor problem~\cite{Calibbi16}.
In a broader class of models, $X$ is considered as an 
axion-like particle (ALP) that acts as a pseudoscalar mediator~\cite{DolanEtAl15}.
Alternatively the introduction of a light, feebly-coupled, spin-1 boson can effectively generate through its axial couplings the phenomenology related to an invisible spin-0 ALP~\cite{Feyet06}.

Searches for $X$ production in the $\KpiX$ decay have the potential to constrain many BSM models.
The $\KpiX$ decay is characterised by an incoming $K^{+}$, an outgoing $\pi^{+}$ and missing energy-momentum, as is the rare $\Kpnn$ decay. 
An interpretation of the NA62 $\Kpnn$ studies using 2017 data~\cite{NA62PNN17} in terms of a search for the $\KpiX$ decay is presented here.
Upper limits are established on $\text{BR}(\KpiX)$ and interpreted in terms of two BSM scenarios. 

\section{Beamline, detector and dataset}
\label{sec:NA62DetectorInfo}
\begin{figure}
  \begin{center}
    \includegraphics[width=1.0\textwidth]{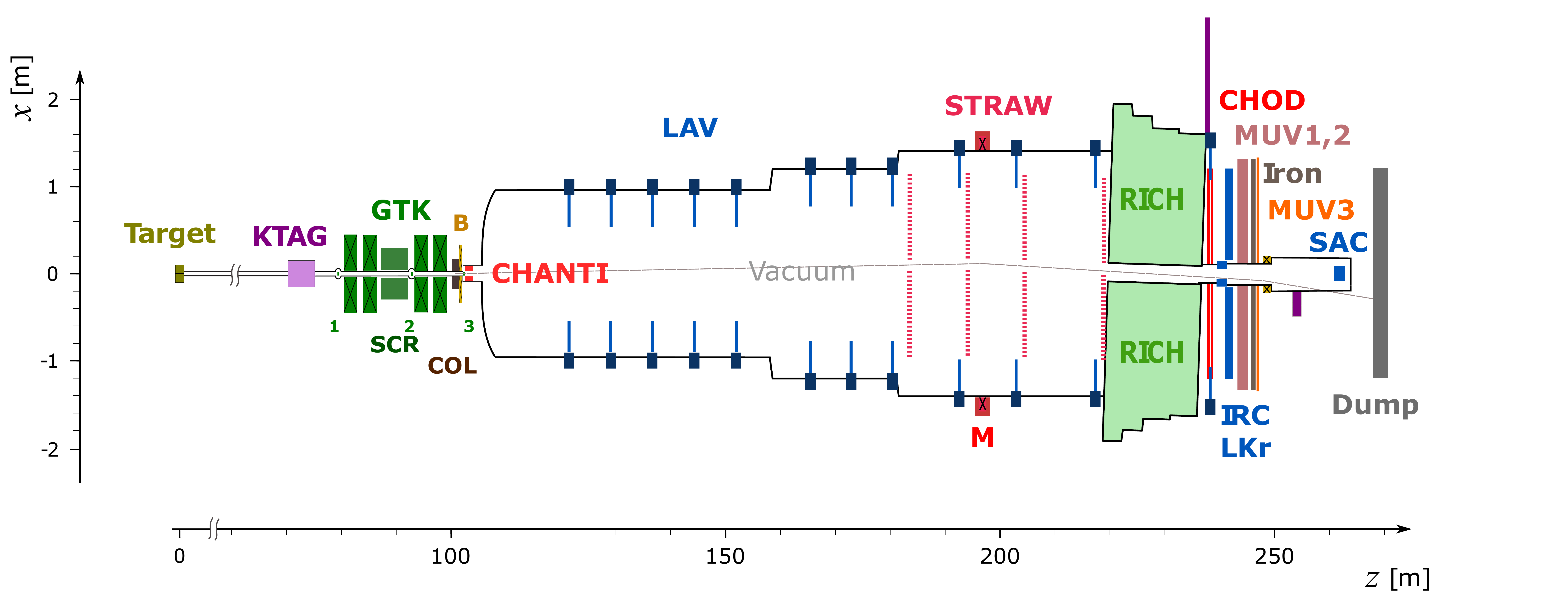}
    \caption{Schematic top view of the NA62 beamline and detector.
    The “CHOD” label indicates both the CHOD and NA48-CHOD hodoscopes described in the text. Also shown is the trajectory of a beam particle in vacuum which crosses all the detector apertures, thus avoiding interactions with material. A dipole magnet between MUV3 and SAC deflects the beam particles out of the SAC acceptance.
    } 
    \label{fig:NA62Detector}
  \end{center}
\vspace{-15pt}
\end{figure}

The NA62 experiment, beamline and detector are described in detail in~\cite{NA62Detector17} and a schematic of the detector is shown in Fig.~\ref{fig:NA62Detector}. 
A right-handed coordinate system, $(x,y,z)$, is defined with the target at the origin and the beam travelling towards positive $z$, the $y$ axis is vertical (positive up) and the $x$-axis is horizontal (positive left).
A $400\,\text{GeV/}c$ proton beam extracted from the CERN Super Proton Synchrotron (SPS) impinges on a beryllium target creating a $75\,\text{GeV/}c$ secondary hadron beam with a $1\%$ rms momentum spread and a composition of $70\%$ pions, $23\%$ protons and $6\%$ kaons. Kaons ($K^{+}$) are positively tagged with $70\,\text{ps}$ timing precision by the KTAG detector, a differential Cherenkov counter filled with nitrogen gas.
The momentum and position of the $K^{+}$ are measured by the GigaTracker (GTK), a spectrometer formed of three silicon pixel tracker stations and a set of four dipole magnets. GTK measurements have momentum, direction and time resolutions of  $0.15\,\text{GeV}/c$, $16\,\mu\text{rad}$ and $100\,\text{ps}$, respectively.
After traversing the GTK magnets, a magnetized scraper used to sweep away muons, and a bending magnet (B), the beam at the FV entrance has a rectangular profile of $52\times24\,\text{mm}^{2}$ and a divergence of $0.11\,\text{mrad}$.

The experiment is designed to study $K^{+}$ decays occurring in the $60\,\text{m}$ fiducial volume (FV) starting $2.6\,\text{m}$ downstream of GTK3 and housed inside a $117\,\text{m}$ long vacuum tank, containing a magnetic spectrometer, and ending at the ring imaging Cherenkov counter (RICH).  
Momentum and position measurements for charged particles produced in $K^{+}$ decays in the FV are provided by the magnetic spectrometer composed of four STRAW tracking stations, two on either side of a dipole magnet (M).
This spectrometer provides a momentum measurement with resolution $\sigma_{p}/p$ of $0.3$--$0.4\%$. The RICH is filled with neon gas at atmospheric pressure and provides particle identification for charged particles, and a time measurement with a precision better than $100\,\text{ps}$. 
Two adjacent scintillator hodoscopes (CHOD and NA48-CHOD), provide time measurements for charged particles with a $200\,\text{ps}$ resolution. 

A system of veto detectors is key to the experiment. Interactions of beam particles in GTK3 are detected by the charged particle anti-counter (CHANTI), formed of six stations of scintillator bar counters. 
Downstream, a photon veto system is used to reject the $K^{+}\rightarrow\pi^{+}\pi^{0}$ background. 
This analysis selects $\pi^{+}$ particles with momenta in the range $15$--$35\,\text{GeV/}c$. This means that a $\pi^{0}$ from the $K^{+}\rightarrow\pi^{+}\pi^{0}$ background has momentum of at least $40\,\text{GeV/}c$ and the subsequent $\pi^{0}\rightarrow\gamma\gamma$ decay, $\text{BR}=98.8\%$, produces two energetic photons which can be detected with high efficiency.
There are twelve large angle veto (LAV) stations positioned to ensure hermetic coverage for photon emission angles of $8.5$--$50\,\text{mrad}$.
The liquid krypton calorimeter (LKr) provides coverage for $1$--$8.5\,\text{mrad}$.
The small angle photon veto (SAV) covers angles below $1\,\text{mrad}$ using two sampling calorimeters of shashlyk design (IRC and SAC). 

Downstream of the LKr are two hadronic sampling calorimeters (MUV1 and MUV2). Together with the LKr, these provide particle identification information through the pattern of energy deposition. Electrons/positrons produce electromagnetic showers that are well-contained in the LKr, which has a depth of 27 radiation lengths. Pions may pass through the LKr without losing all of their energy and can produce a hadronic shower in MUV1 and MUV2. In contrast, muons are minimum ionising particles in the calorimetric system.
The MUV3 detector is positioned downstream of a $0.8\,\text{m}$ iron absorber and consists of a plane of scintillator tiles.
It provides measurements of muons with $400\,\text{ps}$ time resolution.

A two-level trigger system is employed with a hardware level 0 (L0) selection followed by a level 1 (L1) decision made by software algorithms. 
The primary trigger stream of the experiment is dedicated to collection of $\Kpnn$ events and uses information from the CHOD, RICH, LKr, MUV3 at L0~\cite{L0TriggerPaper} and KTAG, LAV, STRAW at L1~\cite{NA62PNN17}.
The NA48-CHOD also provides a $99\%$ efficient minimum bias trigger, used for collection of $K^{+}\rightarrow\pi^{+}\pi^{0}$ events that are used for normalisation. 
The data sample collected in 2017 for the study of the $\Kpnn$ decay is used for this analysis.

\section{Signal selection}
\label{sec:SignalSel}
The observable for the $\KpiX$ search is the reconstructed squared missing mass 
\begin{equation*}
 \mmis = ( P_{K} - P_{\pi})^{2} \,,
\end{equation*}
where $P_{K}$ and $P_{\pi}$ are the $K^{+}$ and $\pi^{+}$ 4-momenta, derived from the measured 3-momenta of the GTK and STRAW tracks under the $K^{+}$ and $\pi^{+}$ mass hypotheses, respectively.
The event selection is identical to that used for the $\Kpnn$ measurement~\cite{NA62PNN17} and is summarised below.

Candidate events must have fewer than three reconstructed STRAW tracks with no negatively charged tracks. 
Only one track can fulfil additional criteria to become a $\pi^{+}$ candidate but, for example, an additional out-of-time halo muon track may exist.
The time assigned to the $\pi^{+}$ candidate is calculated using the mean times measured in the STRAW, NA48-CHOD and RICH weighted by their respective measured resolutions. 
A $\pi^{+}$ candidate track must have momentum in the range $15$--$35\,\text{GeV}/c$ and be within the sensitive regions of the downstream detectors (RICH, CHODs, LKr and MUV1,2,3) with geometrically and time-coincident associated signals recorded in the CHODs, LKr and RICH.

The candidate track must be consistent with the $\pi^{+}$ hypothesis for the RICH reconstructed mass and likelihood. 
The candidate must also satisfy a multivariate classifier based on calorimetric information.
On average, for $15$--$35\,\text{GeV}/c$ tracks, the two methods achieve  $\pi^{+}$ identification efficiencies of $82\%$ and $78\%$, with probabilities of misidentification of $\mu^{+}$ as $\pi^{+}$ of $2.3\times10^{-3}$ and $6.3\times10^{-6}$, respectively. A MUV3 veto condition rejects events with signals geometrically associated with the track within a time window of $7\,\text{ns}$. 
No signals are allowed in any LAV station (or SAV) within $3\,(7)\,\text{ns}$ of the $\pi^{+}$ time.
No LKr clusters are allowed beyond a distance of $100\,\text{mm}$ from the $\pi^{+}$ impact point within cluster-energy dependent time windows of $10$ to $100\,\text{ns}$.
The STRAW, CHODs and LKr are used to veto events with additional activity, including tracks produced by photon interactions upstream of the calorimeters and partially reconstructed multi-track decays.
Overall rejection of $\pi^{0}\rightarrow\gamma\gamma$ decays is achieved with an inefficiency of $1.3\times10^{-8}$.

A $K^{+}$ is tagged upstream by the KTAG if Cherenkov photons are detected within $2\,\text{ns}$ of the $\pi^{+}$ track time in at least five out of its total of eight sectors. 
A GTK track is associated with the $K^{+}$ if its time is within $0.6\,\text{ns}$ of the KTAG time and the closest distance of approach (CDA) to the $\pi^{+}$ track is less than $4\,\text{mm}$. The $K^{+}$/$\pi^{+}$ matching is based on time coincidence and spatial information and has an efficiency of $75\%$. 
The average probability for wrong (accidental) association with pileup GTK tracks is $1.3\%$ ($3.5\%$) when the $K^{+}$ track is (is not) correctly reconstructed.

Upstream backgrounds arise from a combination of early $K^{+}$ decays (upstream of the FV), beam particle interactions in the GTK stations, additional GTK tracks, and large-angle $\pi^{+}$ scattering in the first STRAW station. To minimise such backgrounds, the vertex formed between the selected $K^{+}$ and $\pi^{+}$ tracks must be inside the FV with no additional activity in the CHANTI within $3\,\text{ns}$ of the $\pi^{+}$ candidate time. Additionally, a `box cut' is applied requiring that the projection of the $\pi^{+}$ candidate track back to the final collimator (COL) is outside the area defined by $|x|<100\,\text{mm}$ and $|y|<500\,\text{mm}$.

The $\mmis$ observable is used to discriminate between a peaking two-body $\KpiX$ signal and backgrounds. 
Two signal regions are defined, called region 1 and region 2, to minimise large backgrounds from $K^{+}\rightarrow\pi^{+}\pi^{0}$,  $K^{+}\rightarrow\mu^{+}\nu_{\mu}$ and $K^{+}\rightarrow\pi^{+}\pi^{+}\pi^{-}$ decays.
The reconstructed $\mmis$ for region 1 must be between $0$ and $0.01\,\text{GeV}^{2}/c^{4}$ and that for region 2 between $0.026$ and $0.068\,\text{GeV}^{2}/c^{4}$.
Additional momentum-dependent constraints supplement the definition of the signal regions using alternative squared missing mass variables, constructed either by replacing the GTK measurement of the beam 3-momentum with the average beam momentum and direction, or the STRAW 3-momentum measurement with one measured by the RICH under the $\pi^{+}$ mass hypothesis.
These requirements reject events with incorrect reconstruction of $\mmis$ due to momenta mismeasurements and improve background rejection, but decrease acceptance at the boundaries of the signal regions.

\section{Signal and background models}
\label{sec:SignalAndBkgModels}
\begin{figure}[h]
    \centering
    \includegraphics[width=0.325\textwidth]{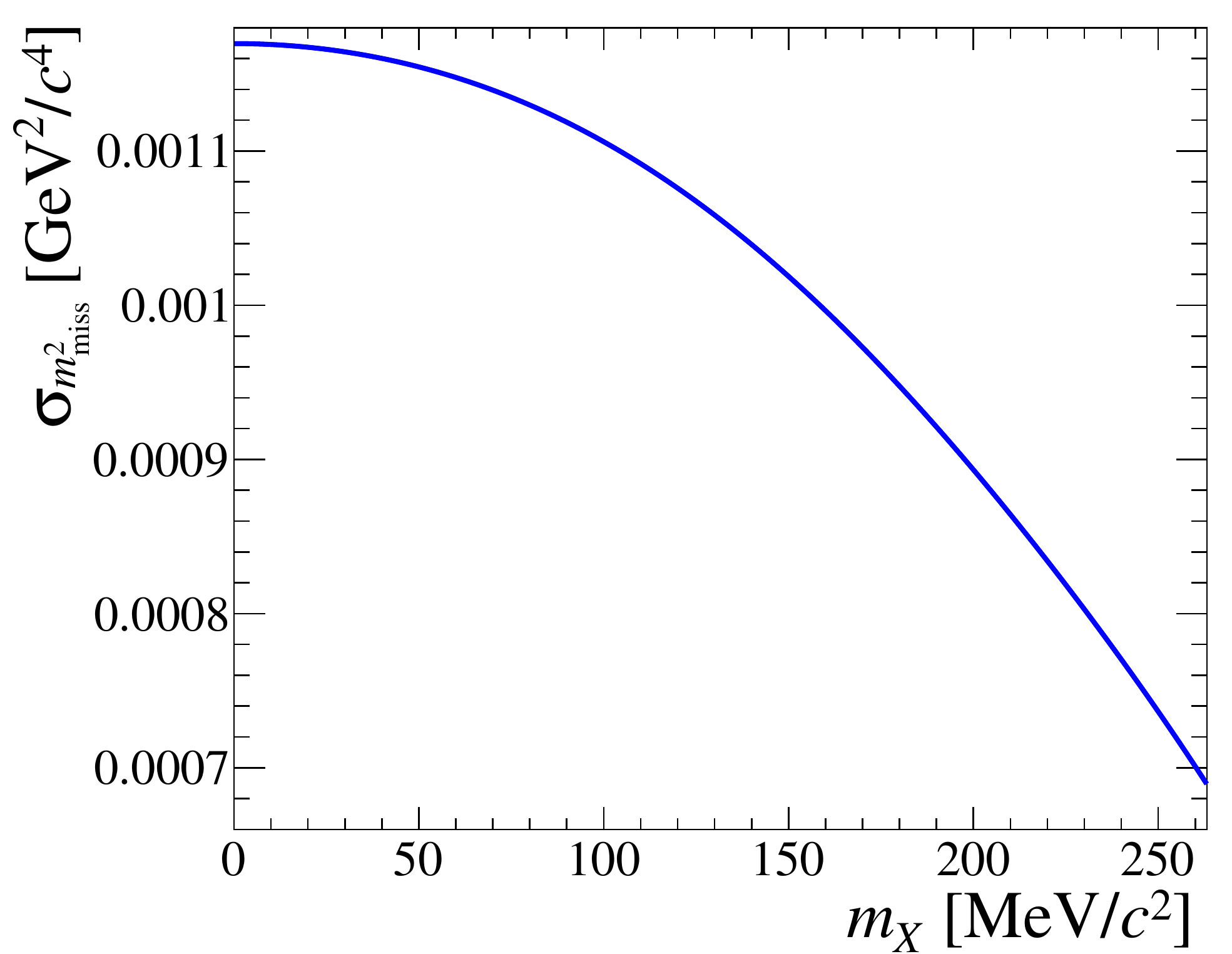}
    \includegraphics[width=0.325\textwidth]{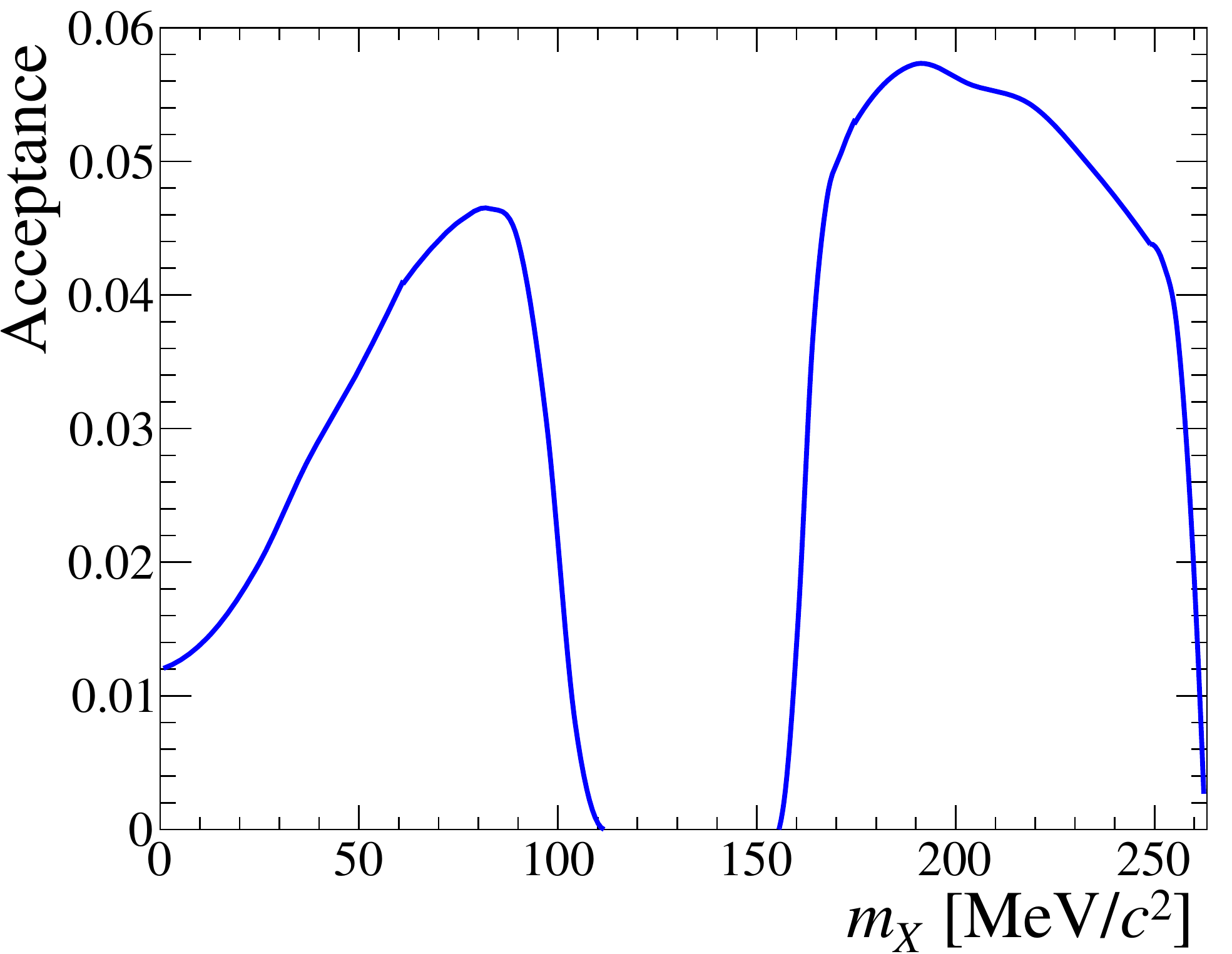}
     \includegraphics[width=0.325\textwidth]{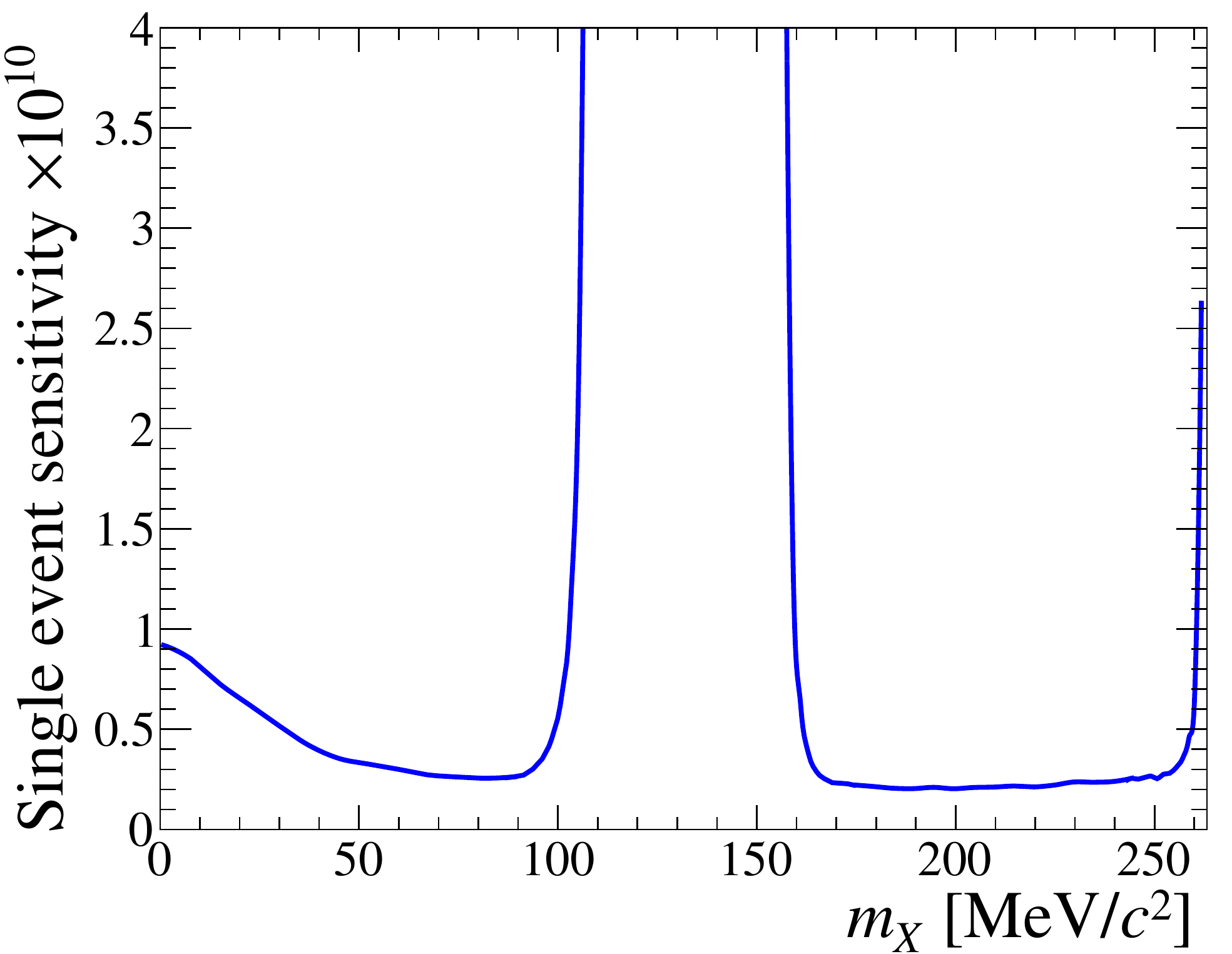} 
     \caption{
       Resolution of the $\mmis$ observable (left), acceptance (centre) and single event sensitivity,
       $\text{BR}_{SES}$, (right) for $K^{+}\rightarrow\pi^{+}X$, as functions of mass hypothesis $m_{X}$.
     }
     \label{fig:ResAccSES}
\end{figure}

\geant-based~\cite{Geant4} Monte Carlo simulations of $\KpiX$ decays are performed with the assumption that $X$ is stable, for $X$ masses covering the search range at $1.4\,\,\text{MeV}/c^{2}$ intervals.
This value corresponds to intervals of the squared missing mass that are always smaller than its resolution. These simulations include decay kinematics, interactions in material, and the responses of the detectors. 
In this study, a scan is performed searching for $\KpiX$ signals with $X$ mass, $m_{X}$, in the ranges $0$--$110\,\text{MeV/}c^{2}$ and $154$--$260\,\text{MeV/}c^{2}$. These $m_{X}$ ranges extend beyond the $\Kpnn$ signal regions because of the resolution of the reconstructed $\mmis$ observable. 
The resolution of $\mmis$, $\sigma_{m^2_{\text{miss}}}$, as a function of simulated $m_{X}$ is shown in Fig.~\ref{fig:ResAccSES} (left).
The reconstructed $\mmis$ resolution for a control sample of selected $K^{+}\rightarrow\pi^{+}\pi^0$ events is found to be $4\%$ better in simulations than in data. The resolution derived from simulations is therefore corrected by increasing it by $4\%$ and a systematic uncertainty of $10\%$ is assigned to the $\mmis$ resolution.
The acceptance for the selection described in section~\ref{sec:SignalSel}, as obtained using simulations, is displayed in Fig.~\ref{fig:ResAccSES} (centre).
The single event sensitivity, $\text{BR}_{SES}$, defined as the branching ratio corresponding to the observation of one signal event, is calculated by following the procedure adopted for the $\Kpnn$ analysis using the $K^{+}\rightarrow\pi^{+}\pi^{0}$ decay for normalisation~\cite{NA62PNN17}; the resulting values are shown in Fig.~\ref{fig:ResAccSES} (right).
The uncertainty of $BR_{SES}$ is $10\%$ and is mainly systematic. The largest contributions to this uncertainty are associated with the trigger efficiency, signal and normalisation reconstruction and selection efficiencies~\cite{NA62PNN17}, and differences between $K^{+}\rightarrow\pi^{+}\nu\bar{\nu}$ and $\KpiX$ kinematics.

The sensitivity for low $X$ masses is limited by the $\Kpnn$ signal region definition $\mmis>0$, which is necessary to suppress the background from $K^{+}\rightarrow\mu^{+}\nu_{\mu}$ decays. 
This effect reduces the acceptance by half for $m_{X}=0$, and equivalently at each signal region boundary (Fig.~\ref{fig:ResAccSES} centre).

The acceptance for $X$ with finite lifetime, $\tau_{X}$ and $m_{X}\ne 0$, is computed under the following assumptions: $X$ decays only to visible SM particles; decays upstream of MUV3 are detected with $100\%$ efficiency.
The efficiency is
$99.9\%$, and the uncertainty in this quantity is included in the systematic
uncertainty.
The acceptance for a set of $\tau_{X}$ values is calculated by weighting simulated events by the probability that $X$ does not decay upstream of MUV3. 
The acceptance increases as a function of lifetime reaching a plateau for $\tau_{X}>10\,\text{ns}$.
For $m_{X}<20\,\text{MeV/}c^{2}$, losses of acceptance at lower lifetimes are compensated by the increase in the Lorentz factor.

\begin{figure}
        \centering
        \vspace{-10pt}
        \includegraphics[width=1.\textwidth]{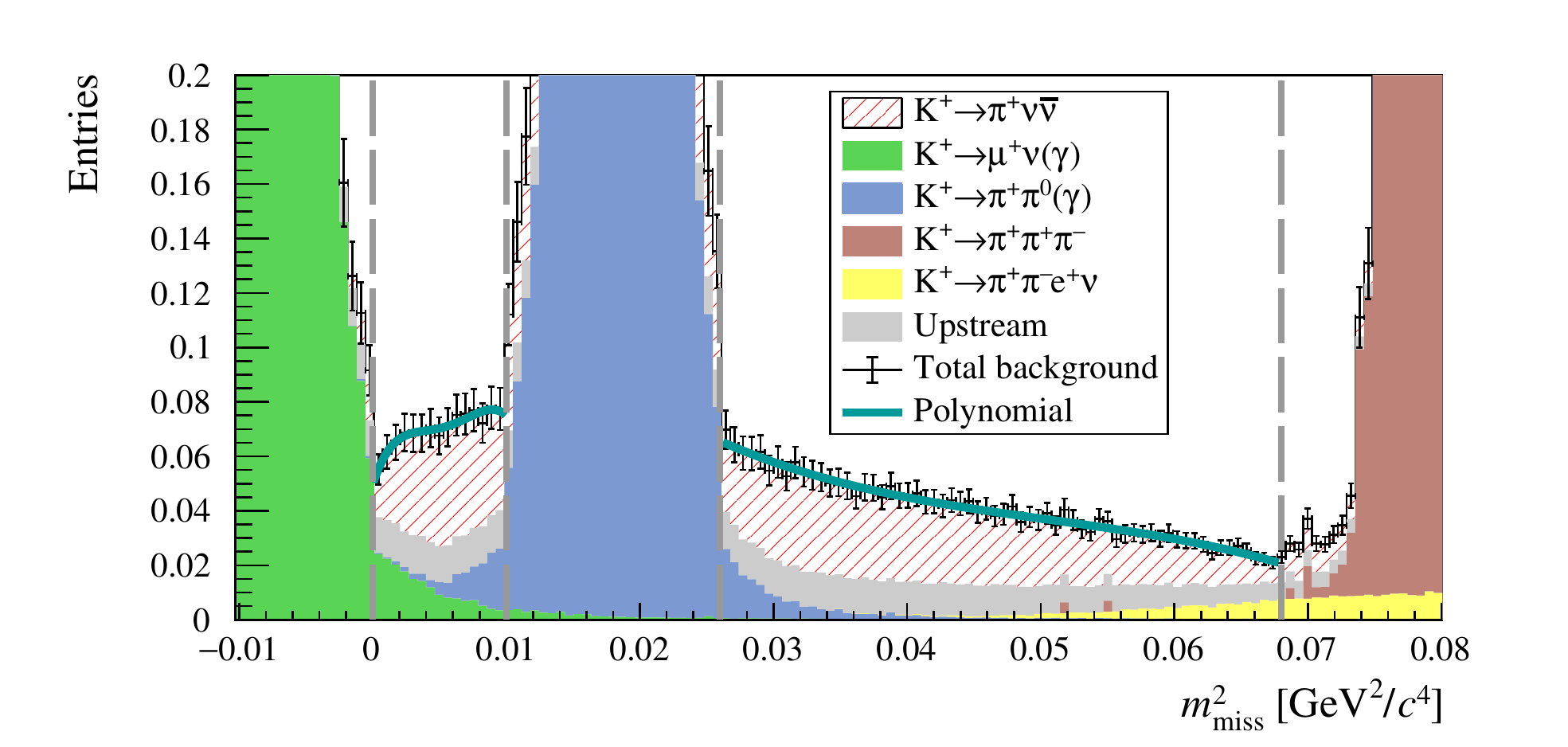}
        \vspace{-10pt}
        \caption{
          Distributions of the expected reconstructed squared missing mass, $\mmis$, for background processes,
          obtained from simulations and data-driven procedures, displayed as stacked histograms with bin width $0.00067\,\text{GeV}^{2}/c^{4}$.
          In each signal region, the polynomial function used to describe the total background is shown.}
        \label{fig:BkgShape}
\end{figure}

The background contributions for the $\KpiX$ search are the same as for the $\Kpnn$ analysis with the addition of the $\Kpnn$ decay itself, which becomes the dominant background. The SM description of the $\Kpnn$ decay is assumed. 
The total expected background and the reconstructed $\mmis$ distributions for each component are obtained from auxiliary measurements, as described in~\cite{NA62PNN17}. The resulting numbers of background events in the signal regions are summarised in Table~\ref{tab:bkg}.
The contributions from kaon decays other than $\Kpnn$ are grouped in the row \textit{other $K^+$ decays},
and their distribution in $\mmis$ is known with good accuracy.
For the \textit{upstream background}, an additional systematic uncertainty of $30\%$ is included, to account for the uncertainty in the estimation of its distribution in $\mmis$ resulting from the limited size of the control sample used for the auxiliary measurements.
The total background is described, as a function of the reconstructed $\mmis$, by fitting polynomial functions to the expectations
in signal regions 1 and 2, as shown in Fig.~\ref{fig:BkgShape}.

\begin{table}[ht]
    \caption{Summary of the predicted numbers of background events in the signal regions and the observed events. The statistical uncertainty for SM $\Kpnn$ is negligible and the external uncertainty arises from the uncertainty of the SM $\Kpnn$ branching ratio. 
    }
    \centerline{
    \resizebox{1.0\textwidth}{!}{
        \begin{tabular}{|l|l|l|l|}
            \hline
                & Region 1 & Region 2 \\
            \hline
                $\Kpnn$ (SM) & $0.55$ $\phantom{\pm0.00_{stat}}$ $\pm\,0.07_{syst}\pm0.13_{ext}$ & $1.61$ $\phantom{\pm0.00_{stat}}$ $\pm\,0.11_{syst}\pm0.22_{ext}$  \\
             Upstream background & $0.21\pm0.12_{stat}\pm0.12_{syst}$ & $0.68\pm0.21_{stat}\pm0.26_{syst}$ \\
                Other $K^{+}$ decays & $0.26\pm0.04_{stat}\pm0.05_{syst}$ & $0.31\pm0.04_{stat}\pm0.06_{syst}$ \\
            \hline
            Total background & $1.02\pm0.13_{stat}\pm0.15_{syst}\pm0.13_{ext}$ & $2.60\pm0.21_{stat}\pm0.28_{syst}\pm0.22_{ext}$ \\
            \hline
                Observed events & $0$ & $2$ \\
            \hline
        \end{tabular}
    }
    }
    \label{tab:bkg}
\end{table}

\section{Statistical analysis} 
The search procedure involves a fully frequentist hypothesis-test using a shape analysis with observable $\mmis$ and an unbinned profile likelihood ratio test statistic. 
Each $X$ mass hypothesis is treated independently.
The parameter of interest, $\text{BR}(\KpiX)$, is related to the expected number of signal events, $n_{S}$, by $\text{BR}(\KpiX)=n_{S}\times\text{BR}_{SES}$. 

The likelihood function has the form:
\begin{equation*}
  \mathcal{L} = 
 \frac{(n_{tot})^n e^{-n_{tot}}}{n!}
  \times \prod_{j}^{n} \Bigg[ \frac{n_B}{n_{tot}} f_{B} \Big(\mmis,_{j}\Big) + \frac{n_{S}}{n_{tot}}  f_{S} \Big(\mmis,_{j}|\mu_{X},\sigma_{X}\Big) \Bigg] \times
  \prod^{N_{\text{nuis}}}_{i} \mathcal{C}_{i}(p_{\text{meas}}^{i}|p_{\text{nuis}}^{i}) \,
\end{equation*}
where $n$ is the observed number of events, $n_{tot}=n_B+n_S$ and
$n_{B}$ is the expected number of background events; $f_{B}(\mmis)$ is a polynomial function of $\mmis$ normalised to unity which describes the total background in the signal region relevant for a certain mass hypothesis $m_X$; and $f_{S}(\mmis|\mu_{X},\sigma_{X})$ is the Gaussian function, normalised to unity, with parameters $\mu_{X}$ and $\sigma_{X}$ obtained from a fit to the distribution of the reconstructed simulated events.
Index $j$ runs over the $n$ observed events and their reconstructed $\mmis$ are denoted $\mmis,_{j}$.
The $N_{\text{nuis}}$ nuisance parameters considered, $p_{\text{nuis}}^{i}$, are $n_{B},\,\text{BR}_{SES},\,\mu_{X},\,\sigma_{X}$,
and are estimated by auxiliary measurements.
These estimations, $p_{\text{meas}}^{i}=\hat{n}_{B},\,\hat{\text{BR}}_{SES},\,\hat{\mu}_{X},\,\hat{\sigma}_{X}$, are treated as global observables ~\cite{LogLikelihoodStats}.
The constraint terms, $\mathcal{C}_{i}(p_{\text{meas}}^{i}|p_{\text{nuis}}^{i})$, are the probability density functions describing the distribution of each nuisance parameter.
The constraint term for $n_{B}$ is a Poisson distribution with mean value
$(\hat{n}_{B}/\sigma_{B})^{2}$ where $\hat{n}_{B}$ and $\sigma_{B}$ are the central value and uncertainty of the background expectation~\cite{CousinsEtAl07}.
The constraint term for $\text{BR}_{SES}$ is a log-normal function with parameters corresponding to a relative uncertainty of $10\%$.
A Gaussian constraint term is used for $\mu_{X}$, with relative uncertainty depending on the mass hypothesis $m_{X}$.   
A log-normal constraint term is used for $\sigma_{X}$, with the mean corresponding to the estimated value after the $4\%$ correction (described in Section~\ref{sec:SignalAndBkgModels}), and relative uncertainty of 10\%.
The normalised polynomial functions, describing the background distribution in $\mmis$, are considered to be known exactly.

For each mass hypothesis the fully frequentist test is performed
according to the CLs method~\cite{Read02} to exclude the presence of a signal with $90\%$ confidence level (CL)
for the observed data. 
A cross-check was performed, using single bin counting experiments in windows of width equal to four times $\sigma_{m^2_{\text miss}}$ around each mass hypothesis, with a hybrid frequentist treatment using a log-likelihood ratio test statistic. A comparable expected sensitivity was obtained.

\section{Results and discussion}
\begin{figure}
    \centering
    \begin{subfigure}[b]{0.49\textwidth}
        \hspace{-20pt} \vspace{-5pt}
        \includegraphics[width=1.14\textwidth]{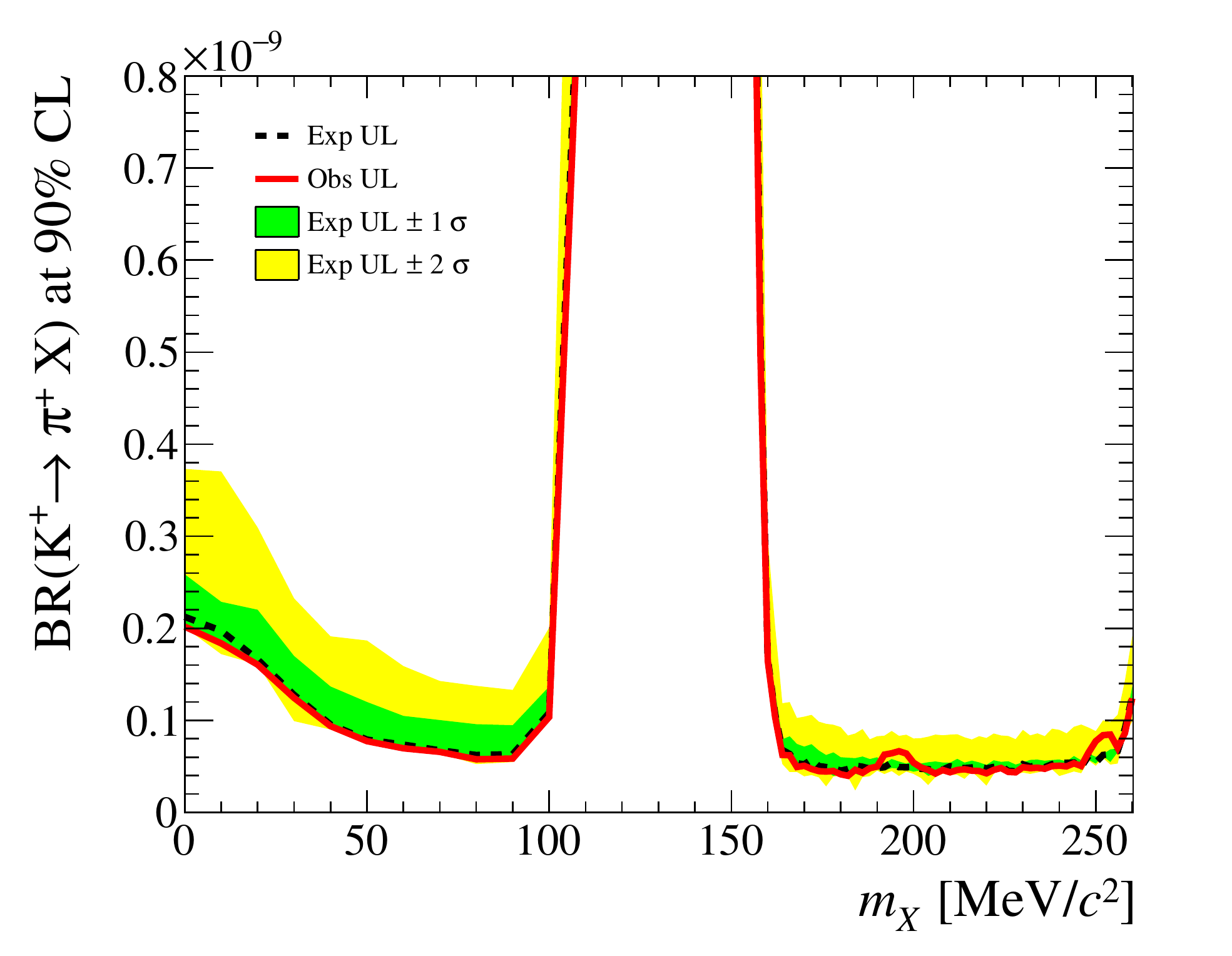}
        \caption*{} 
        \label{fig:NA62BRUL} 
    \end{subfigure}
    \begin{subfigure}[b]{0.49\textwidth}
        \includegraphics[width=1.05\textwidth]{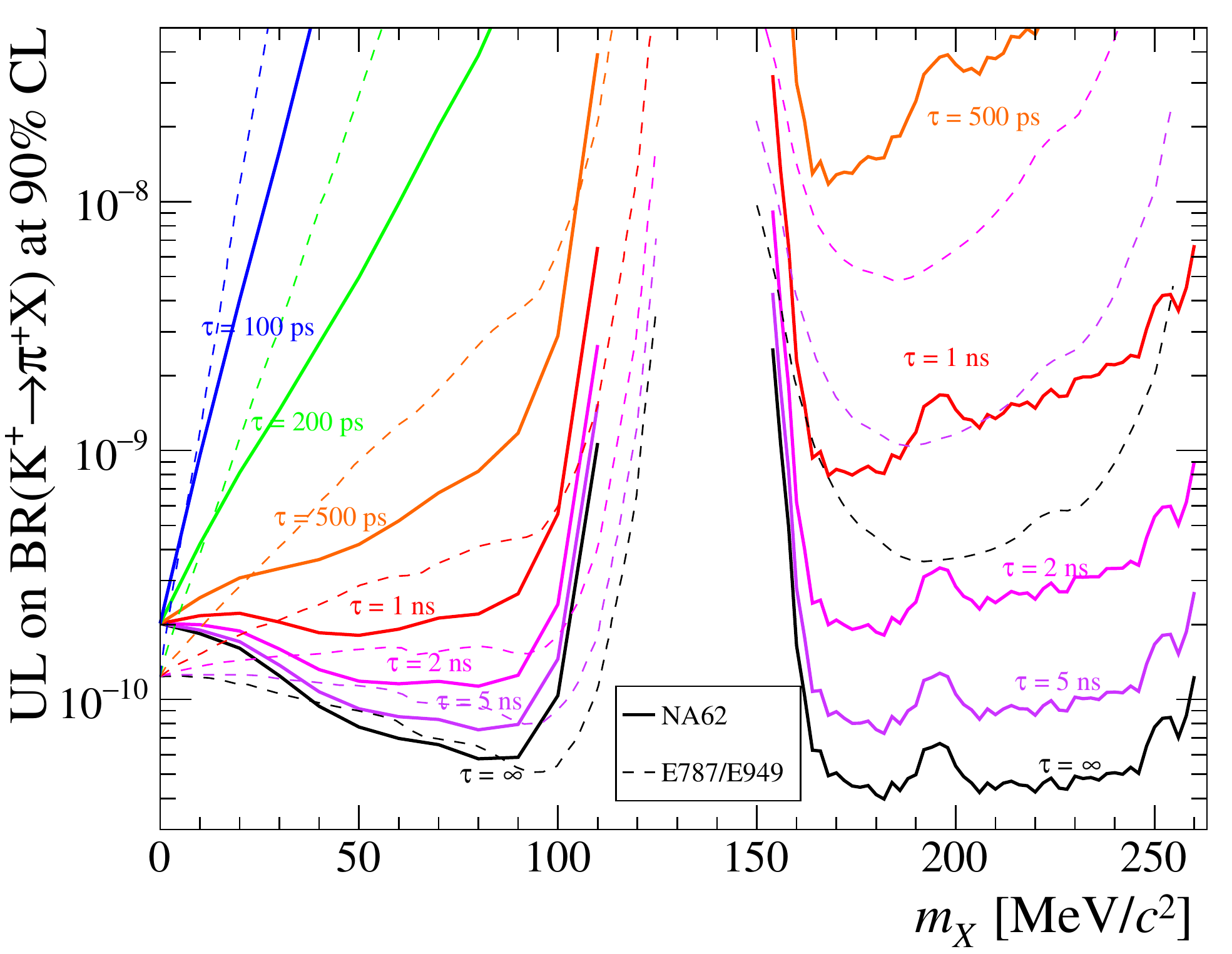}
        \caption*{}
        \label{fig:BrULlifetimes}
    \end{subfigure}
    \vspace{-21pt}
    \caption{
    Left: upper limits on $\text{BR}(\KpiX)$ for each mass hypothesis, $m_{X}$,  tested. Right: model-independent observed upper limits as functions of the mass and lifetime assumed for $X$ for NA62 (this work, solid lines) and E787/E949~\cite{BNL09} (dashed lines).
    } 
    \label{fig:KpiXBrUL}
\end{figure}

Two candidate $\KpiX$ events are observed~\cite{NA62PNN17} at reconstructed $m_{\text{miss}}$ values of $196$ and $252\,\text{MeV}/c^{2}$. 
Upper limits are established on $\text{BR}(\KpiX)$ at $90\%$ CL for each $X$ mass hypothesis:
expected and observed upper limits, assuming stable or invisibly decaying $X$, are displayed in Fig.~\ref{fig:KpiXBrUL} (left). The observed upper limits are compared to the previous results from the E787/E949 experiments~\cite{BNL09} in Fig.~\ref{fig:KpiXBrUL} (right), as a function of $m_{X}$ and for different values of $\tau_{X}$, assuming $X$ decays to visible SM particles. 
The strongest limits of $5\times10^{-11}$ are obtained at large $X$ masses ($160$--$250\,\text{MeV}/c^{2}$) and long $X$ lifetimes ($>5\,\text{ns}$).
Under the assumption of stable or invisibly decaying $X$ these upper limits improve  by a factor of $\mathcal{O}(10)$ in signal region 2, and are competitive in region 1. For unstable $X$, assuming decays only to visible SM particles, the same pattern holds in general. However, in region 1 the limits obtained improve across an increasingly large range of mass hypotheses as the assumed lifetime becomes shorter.
Despite differences in experimental set-up between E787/E949 (stopped $K^{+}$ decay-at-rest) and NA62 (highly boosted $K^{+}$ decay-in-flight), the two results exhibit similar dependence on $\tau_{X}$. 
This is because the ratios of the Lorentz factor for the $X$ particle to the decay length are similar in the two experiments.

\begin{figure}
    \centering
    \begin{subfigure}[b]{0.49\textwidth}
        \hspace{-20pt}
        \includegraphics[width=1.05\textwidth]{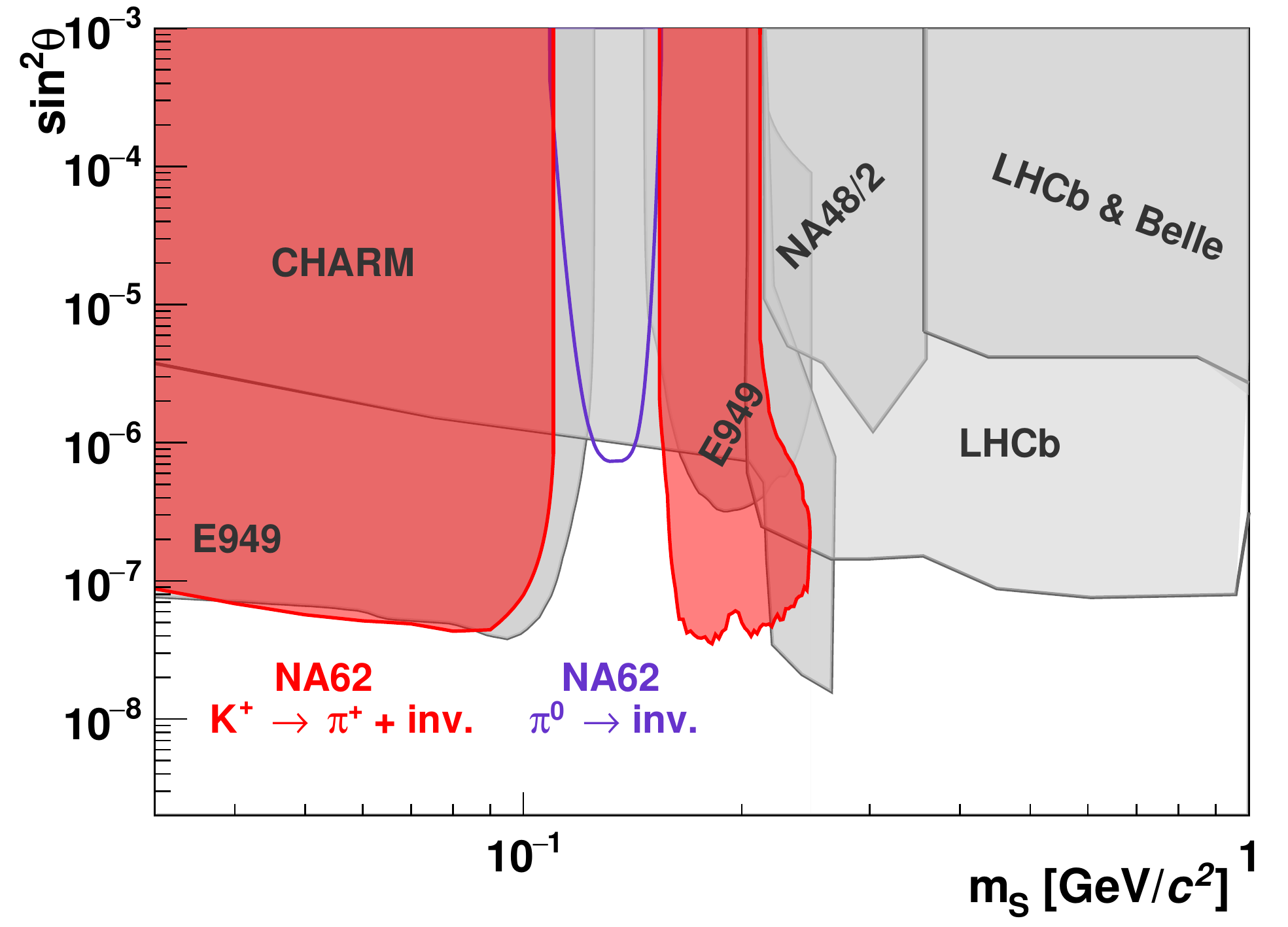}
        \caption*{} 
        \label{fig:KpiX_ScalarBC4_Visible} 
    \end{subfigure}
    \begin{subfigure}[b]{0.49\textwidth}
        \includegraphics[width=1.05\textwidth]{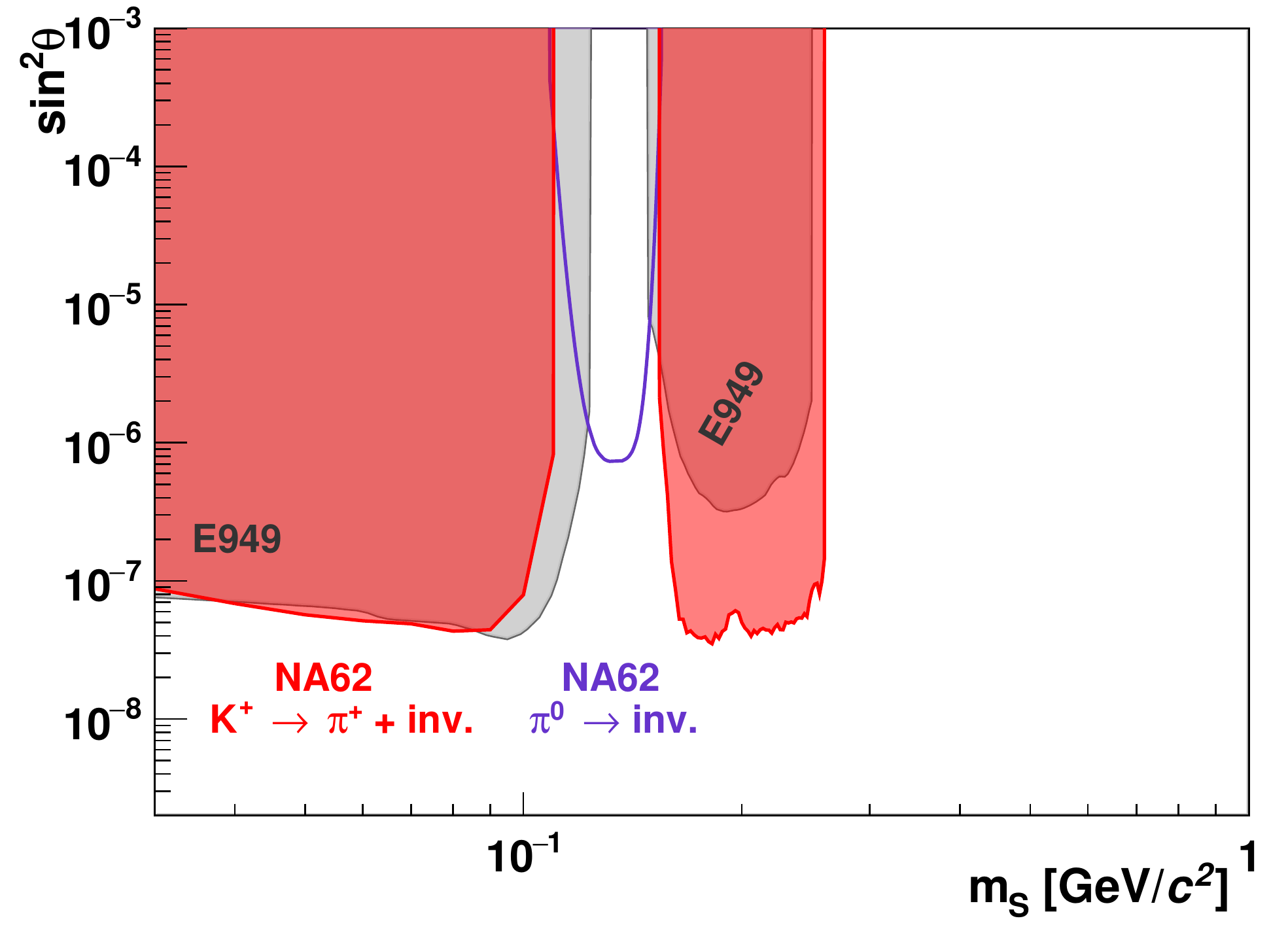}
        \caption*{}
        \label{fig:KpiX_ScalarBC4_Invisible}
    \end{subfigure}
    \vspace{-17pt}
    \caption{
    Excluded regions of the parameter space $(m_{S},\sin^{2}\theta)$ for a dark scalar, $S$, of the BC4 model~\cite{PBC19} decaying only (left) to visible SM particles as in the BC4 model and (right) invisibly. 
    The exclusion bound from the present search for the decay $K^{+}\rightarrow\pi^{+}S$ is labelled as ``$K^{+}\rightarrow\pi^{+}+\text{inv.}$'' and is shaded in red. 
    In the $\pi^{0}$ mass region the independent NA62 search for $\pi^{0}\rightarrow\text{invisible}$ decays~\cite{NA62pi0inv} provides constraints, shown in purple.
    Other bounds, shown in grey, are derived from the experiments E949~\cite{BNL09}, CHARM~\cite{Winkler}, NA48/2~\cite{NA48Kpimumu}, LHCb~\cite{LHCbBKmumu_a,LHCbBKmumu_b} and Belle~\cite{BelleBKmumu}.
    } 
    \label{fig:KpiX_ScalarBC4}
\end{figure}

In a Higgs portal model with a dark sector scalar mixing with the Higgs boson, $X$ production and decay are driven by the mixing parameter $\sin^{2}\theta$ (model BC4~\cite{PBC19,Winkler}).
This gives rise to $K^{+}\rightarrow\pi^{+}X$ decays with branching ratio proportional to $\sin^{2}\theta$. 
The constraints derived on $\sin^{2}\theta$ from this search, alongside results from other studies, are shown in Fig.~\ref{fig:KpiX_ScalarBC4}.

\begin{figure}
    \centering
    \begin{subfigure}[b]{0.49\textwidth}
        \hspace{-20pt}
        \includegraphics[width=1.05\textwidth]{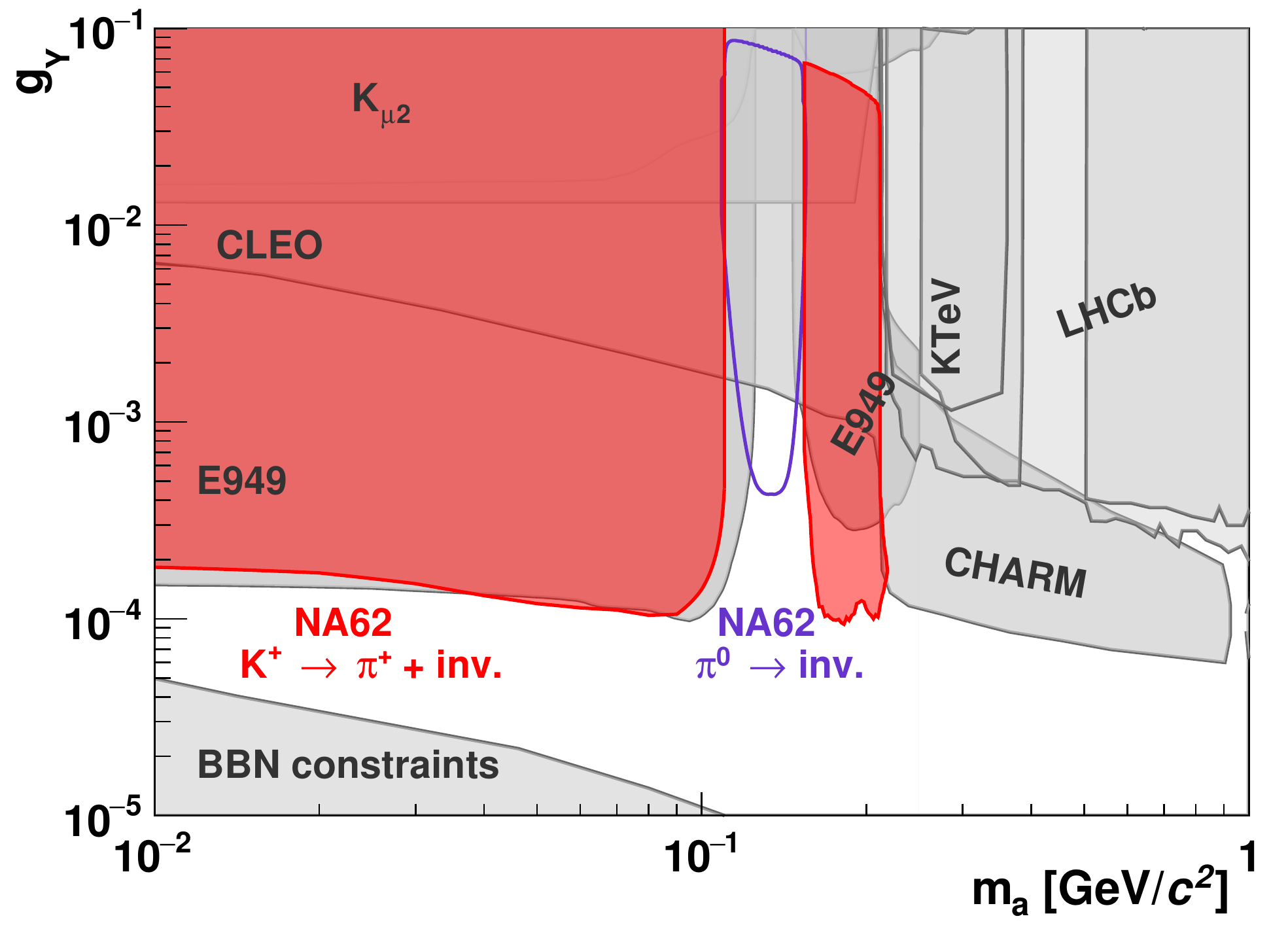}
        \caption*{} 
        \label{fig:KpiX_ALPBC10_Visible} 
    \end{subfigure}
    \begin{subfigure}[b]{0.49\textwidth}
        \includegraphics[width=1.05\textwidth]{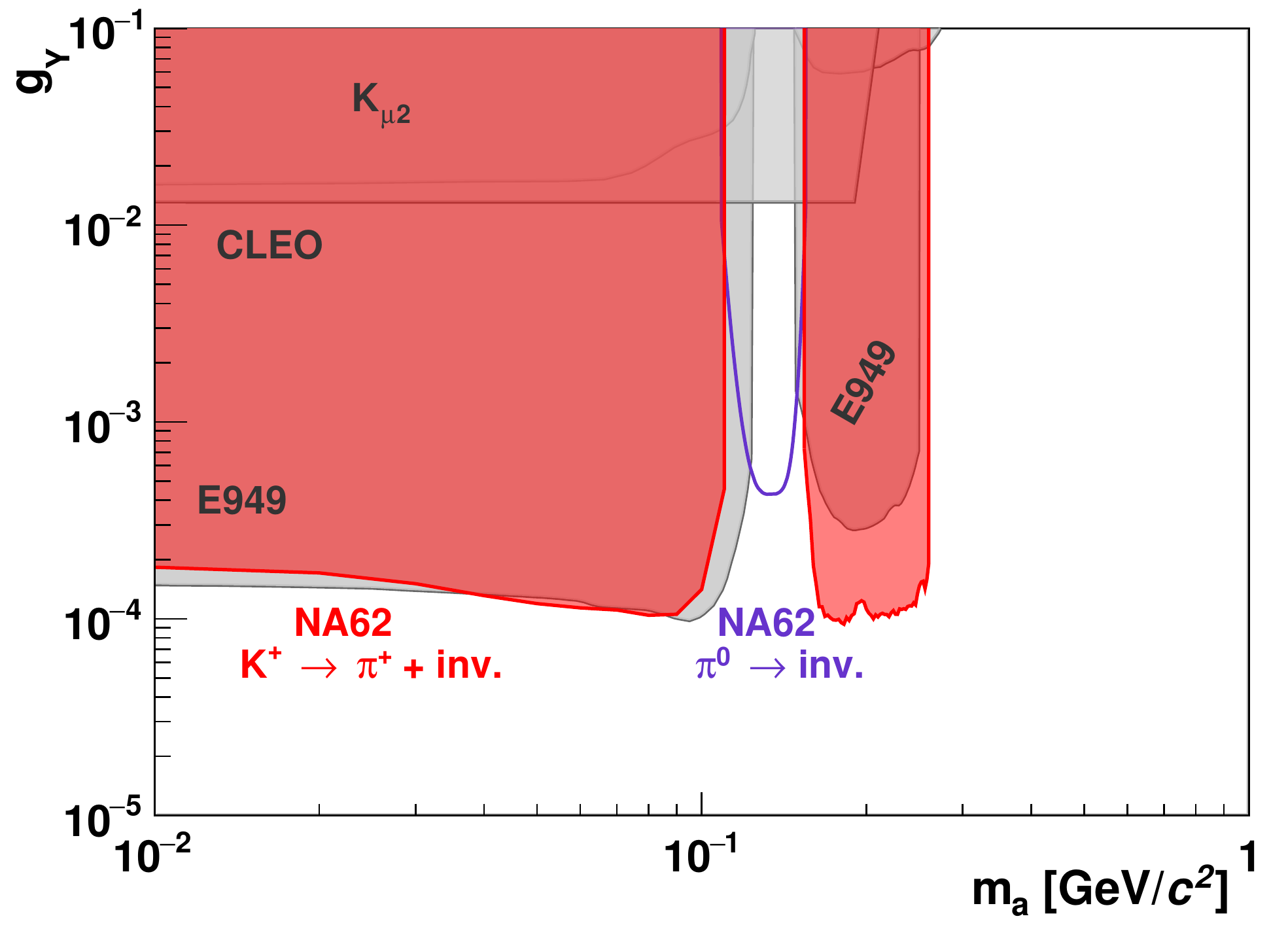}
        \caption*{}
        \label{fig:KpiX_ALPBC10_Invisible}
    \end{subfigure}
    \vspace{-17pt}
    \caption{
     Excluded regions of the parameter space $(m_{a},g_{Y})$ for an ALP, $a$, of the BC10 model~\cite{PBC19} decaying only (left) to visible particles and (right) invisibly.
     The exclusion bound from the present search for the decay $K^{+}\rightarrow\pi^{+}a$ is labelled as ``$K^{+}\rightarrow\pi^{+}+\text{inv.}$'' and is shaded in red. 
     In the $\pi^{0}$ mass region the independent NA62 search for $\pi^{0}\rightarrow\text{invisible}$ decays~\cite{NA62pi0inv} provides constraints, shown in purple. 
     Other bounds, shown in grey, are derived from the experiments E949~\cite{BNL09}, $K_{\mu2}$~\cite{Kmu2ALP}, CLEO~\cite{CLEO_ALP}, CHARM~\cite{CHARM_ALP}, KTeV~\cite{KTeV}, LHCb~\cite{LHCbBKmumu_a,LHCbBKmumu_b} and from Big Bang nucleosynthesis (BBN)~\cite{PBC19}.
    } 
    \label{fig:KpiX_ALPBC10}
\end{figure}

In a scenario where $X$ is an ALP with couplings proportional to SM Yukawa couplings (model BC10~\cite{PBC19,DolanEtAl15}) the $K^{+}\rightarrow\pi^{+}X$ decay occurs with a branching ratio proportional to the square of the coupling constant $g_{Y}$. The constraints on $g_{Y}$ derived from this  and other searches are shown in Fig.~\ref{fig:KpiX_ALPBC10}. 

If $X$ decays only to invisible particles, such as dark matter, bounds on the coupling parameter ($\sin^{2}\theta$ or $g_{Y}$ for the scalar and ALP models, respectively) are directly derived from its relationship with the branching ratio, with results shown in the right-hand panels of Figs.~\ref{fig:KpiX_ScalarBC4} and~\ref{fig:KpiX_ALPBC10}.
If $X$ decays only to visible SM particles, $\tau_{X}$ is inversely proportional to the coupling parameters~\cite{DolanEtAl15,Winkler}, limiting the reach of this analysis for large coupling because of lower acceptance for shorter lifetimes. 
The $X\rightarrow e^{+}e^{-}$ decays dominate the visible decay width up to the di-muon threshold beyond which an additional channel opens and $\tau_{X}$ decreases, limiting the sensitivity of this search.
The model-dependent relationship between the lifetime and coupling therefore determines the shape of the exclusion regions shown in the left-hand panels of Figs.~\ref{fig:KpiX_ScalarBC4} and~\ref{fig:KpiX_ALPBC10}.

\section{Conclusions}
A search for the $\KpiX$ decay, where $X$ is a long-lived feebly interacting particle, is performed through an interpretation of the $\Kpnn$ analysis of data collected in 2017 by the NA62 experiment at CERN.
Two candidate $\KpiX$ events are observed, in agreement with the expected background. Upper limits on $\text{BR}(\KpiX)$ are established at $90\%$ CL, with the strongest limits of $5\times10^{-11}$ at large $X$ masses ($160$--$250\,\text{MeV}/c^{2}$) and long $X$ lifetimes ($>5\,\text{ns}$), improving on current results by up to a factor of $\mathcal{O}(10)$. 
An interpretation of these results to constrain BSM models is presented in scenarios where $X$ is a dark scalar mixing with the Higgs boson or is an ALP with couplings to fermions. 

\clearpage 
\section*{Acknowledgements}
It is a pleasure to express our appreciation to the staff of the CERN laboratory and the technical
staff of the participating laboratories and universities for their efforts in the operation of the
experiment and data processing.

The cost of the experiment and its auxiliary systems was supported by the funding agencies of 
the Collaboration Institutes. We are particularly indebted to: 
F.R.S.-FNRS (Fonds de la Recherche Scientifique - FNRS), Belgium;
BMES (Ministry of Education, Youth and Science), Bulgaria;
NSERC (Natural Sciences and Engineering Research Council), funding SAPPJ-2018-0017 Canada;
NRC (National Research Council) contribution to TRIUMF, Canada;
MEYS (Ministry of Education, Youth and Sports),  Czech Republic;
BMBF (Bundesministerium f\"{u}r Bildung und Forschung) contracts 05H12UM5, 05H15UMCNA and 05H18UMCNA, Germany;
INFN  (Istituto Nazionale di Fisica Nucleare),  Italy;
MIUR (Ministero dell'Istruzione, dell'Univer\-sit\`a e della Ricerca),  Italy;
CONACyT  (Consejo Nacional de Ciencia y Tecnolog\'{i}a),  Mexico;
IFA (Institute of Atomic Physics) Romanian CERN-RO No.1/16.03.2016 and Nucleus Programme PN 19 06 01 04,  Romania;
INR-RAS (Institute for Nuclear Research of the Russian Academy of Sciences), Moscow, Russia; 
JINR (Joint Institute for Nuclear Research), Dubna, Russia; 
NRC (National Research Center)  ``Kurchatov Institute'' and MESRF (Ministry of Education and Science of the Russian Federation), Russia; 
MESRS  (Ministry of Education, Science, Research and Sport), Slovakia; 
CERN (European Organization for Nuclear Research), Switzerland; 
STFC (Science and Technology Facilities Council), United Kingdom;
NSF (National Science Foundation) Award Numbers 1506088 and 1806430,  U.S.A.;
ERC (European Research Council)  ``UniversaLepto'' advanced grant 268062, ``KaonLepton'' starting grant 336581, Europe.

Individuals have received support from:
Charles University Research Center (UNCE/SCI/ 013), Czech Republic;
Ministry of Education, Universities and Research (MIUR  ``Futuro in ricerca 2012''  grant RBFR12JF2Z, Project GAP), Italy;
Russian Foundation for Basic Research  (RFBR grants 18-32-00072, 18-32-00245), Russia; 
Russian Science Foundation (RSF 19-72-10096), Russia;
the Royal Society  (grants UF100308, UF0758946), United Kingdom;
STFC (Rutherford fellowships ST/J00412X/1, ST/M005798/1), United Kingdom;
ERC (grants 268062,  336581 and  starting grant 802836 ``AxScale'');
EU Horizon 2020 (Marie Sk\l{}odowska-Curie grants 701386, 842407, 893101).

\end{document}